# Some new results concerning the QFT vacuum in the Heisenberg picture

Dan Solomon
Rauland-Borg Corporation
3450 W. Oakton
Skokie, IL 60077  USA

Email: **dan.solomon@rauland.com**
(Feb. 17,  2007)

**Abstract**

It has recently been shown [1] that in Dirac's hole theory the vacuum state is not the minimum energy state but that there exist quantum states with less energy than that of the vacuum state.  In this paper we extend this discussion to quantum field theory (QFT) in the Heisenberg picture and consider the question of whether or not the vacuum in QFT is the state of minimum energy.  It will be shown that for a "simple" field theory, consisting of a quantized fermion field interacting with a classical electric field in 1-1D space-time, there exist quantum states with less energy than that of the vacuum state.

PACS Nos: 11.10-z , 03.65-w

## 1. Introduction

In quantum theory it is generally assumed that the vacuum state is the minimum energy state. However it has been shown that this is not the case for Dirac's hole theory. In Ref. [1] Dirac's Hole theory was examined in 1-1D space-time. It was shown that the vacuum state was not the state of minimum energy and that there existed quantum states with less energy than that of the vacuum state. In this paper we will apply the analysis of Ref. [1] to quantum field theory (QFT) in the Heisenberg picture. We will examine the question of whether or not the unperturbed vacuum state in QFT is the state of minimum energy. We will consider a "simple" field theory in 1-1D space-time consisting of a quantized fermion field interacting with an unquantized, classical electric field. It will be shown that, for this situation, there exist quantum states with less energy than that of the vacuum state.

In the Heisenberg picture the state vectors $|\Omega\rangle$ are constant in time and the time dependence of the quantum state is carried by the field operators $\hat{\psi}(z,t)$ where $z$ is used to represent the space dimension. A quantum state is specified by the state vector $|\Omega\rangle$ and field operator $\hat{\psi}(z,t)$. We will write this as the pair $(\psi, |\Omega\rangle)$. The Dirac Hamiltonian in the presence of a classical electromagnetic field is given by,

$$\hat{H}(\hat{\psi}) = \int \hat{\psi}^\dagger H \hat{\psi} \, dz \tag{1}$$

where,

$$H = H_0 + qV(z,t) \tag{2}$$

and where $V(z,t)$ is an external electric potential, q is a coupling constant, and $H_0$ is given by,

$$H_0 = \left(-i\sigma_x \frac{\partial}{\partial z} + m\sigma_z\right) \tag{3}$$

where $\sigma_x$ and $\sigma_z$ are the usual Pauli matrices.

In the Heisenberg picture the evolution of the field operator is given by,

$$\frac{\partial \hat{\psi}(z,t)}{\partial t} = i\left[\hat{H}(\hat{\psi}(z,t)), \hat{\psi}(z,t)\right] \tag{4}$$

The field operator obeys the equal time anti-commutator relationship,

$$\hat{\psi}^\dagger_\alpha(z,t)\hat{\psi}_\beta(z',t) + \hat{\psi}_\beta(z',t)\hat{\psi}^\dagger_\alpha(z,t) = \delta_{\alpha\beta}\delta(z-z') \quad (5)$$

with all other equal time anti-commutators being equal to zero. It can be shown (see Chapt. 9 of [2] or Section 8 of [3]) that when these are used in (4) we obtain,

$$i\frac{\partial \hat{\psi}(z,t)}{\partial t} = H\hat{\psi}(z,t) \quad (6)$$

## 2. The vacuum state

Assume that at some initial time the electric potential is zero and that field operator is in its initial unperturbed state which we designate by $\hat{\psi}^{(0)}(z,t)$. In that case Eq. (6) becomes,

$$i\frac{\partial \hat{\psi}^{(0)}(z,t)}{\partial t} = H_0\hat{\psi}^{(0)}(z,t) \quad (7)$$

In addition to satisfying the above equation the initial field operator must also satisfy (5). This can be achieved by writing,

$$\hat{\psi}^{(0)}(z,t) = \sum_r \left(\hat{b}_r \varphi^{(0)}_{1,r}(z,t) + \hat{d}^\dagger_r \varphi^{(0)}_{-1,r}(z,t)\right); \quad \hat{\psi}^{(0)\dagger}(z,t) = \sum_r \left(\hat{b}^\dagger_r \varphi^{(0)\dagger}_{1,r}(z,t) + \hat{d}_r \varphi^{(0)\dagger}_{-1,r}(z,t)\right) \quad (8)$$

where the $\hat{b}_r$ ($\hat{b}^\dagger_r$) are the destruction(creation) operators for an electron associated with the state $\varphi^{(0)}_{1,r}(z,t)$ and the $\hat{d}_r$ ($\hat{d}^\dagger_r$) are the destruction(creation) operators for a positron associated with the state $\varphi^{(0)}_{-1,r}(z,t)$. They satisfy the anticommutator relationships,

$$\{\hat{d}_j,\hat{d}^\dagger_k\} = \delta_{jk}; \quad \{\hat{b}_j,\hat{b}^\dagger_k\} = \delta_{jk}; \text{ all other anti-commutators are zero} \quad (9)$$

The vacuum state $|0\rangle$ is defined by,

$$\hat{d}_j|0\rangle = \hat{b}_j|0\rangle = 0 \text{ and } \langle 0|\hat{d}^\dagger_j = \langle 0|\hat{b}^\dagger_j = 0 \text{ for all } j \quad (10)$$

The quantities $\varphi^{(0)}_{\lambda,r}(z,t)$ are solutions of Eq. (7). We will assume periodic boundary conditions so that the solutions satisfy $\varphi^{(0)}_{\lambda,r}(z,t) = \varphi^{(0)}_{\lambda,r}(z+L,t)$ where L is the 1-dimensional integration volume. Therefore, we obtain,

$$\varphi^{(0)}_{\lambda,r}(z,t) = \varphi^{(0)}_{\lambda,r}(z)e^{-i\varepsilon^{(0)}_{\lambda,r}t}; \quad \varphi^{(0)}_{\lambda,r}(z) = u_{\lambda,r}e^{ip_r z} \quad (11)$$

where 'r' is an integer, $\lambda = \pm 1$ is the sign of the energy, $p_r = 2\pi r/L$, and where,

$$\varepsilon_{\lambda,r}^{(0)} = \lambda E_r \; ; \; E_r = +\sqrt{p_r^2 + m^2} \; ; \; u_{\lambda,r} = N_{\lambda,r} \begin{pmatrix} 1 \\ p_r \big/ (\lambda E_r + m) \end{pmatrix} ; \; N_{\lambda,r} = \sqrt{\frac{\lambda E_r + m}{2L\lambda E_r}} \tag{12}$$

The quantities $\varphi_{\lambda,r}^{(0)}(z)$ satisfy the relationship,

$$H_0 \varphi_{\lambda,r}^{(0)}(z) = \varepsilon_{\lambda,r}^{(0)} \varphi_{\lambda,r}^{(0)}(z) \tag{13}$$

The $\varphi_{\lambda,r}^{(0)}(z)$ form an orthonormal basis set and satisfy,

$$\int_{-L/2}^{+L/2} \varphi_{\lambda,r}^{(0)\dagger}(z) \varphi_{\lambda',s}^{(0)}(z) dz = \delta_{\lambda\lambda'} \delta_{rs} \tag{14}$$

### 3. The change in energy

If the electric potential is zero then the energy of the quantum state $(\psi, |\Omega\rangle)$ is given by,

$$\xi(\hat{\psi}, |\Omega\rangle) = \langle \Omega | \hat{H}_0(\hat{\psi}) | \Omega \rangle = \langle \Omega | \int \hat{\psi}^\dagger H_0 \hat{\psi} dz | \Omega \rangle \tag{15}$$

Next we want to consider the following problem. Start with the system in the initial unperturbed vacuum state. In this case the field operator is $\hat{\psi}^{(0)}(z,t)$ and the state vector is $|0\rangle$. The question we want to address is whether or not the unperturbed vacuum state $(\hat{\psi}^{(0)}(z,t), |0\rangle)$ is the state of minimum energy or do there exist states with less energy than that of the vacuum state? In order to address this problem consider the change in the energy of the vacuum state due to an interaction with an external electric potential which is applied at some time $t > t_0$ and then removed at some later time $t_1$ so that,

$$V(z,t) = 0 \text{ for } t \le t_0; \; V(z,t) \ne 0 \text{ for } t_0 < t < t_1; \; V(z,t) = 0 \text{ for } t \ge t_1 \tag{16}$$

Under the action of the electric potential the system evolves from the initial vacuum state $(\hat{\psi}^{(0)}(z,t_0), |0\rangle)$, at $t_0$, to the final state $(\hat{\psi}(z,t_1), |0\rangle)$, at $t_1$. The change in the energy is given by,

$$\Delta \xi_{vac} = \xi(\hat{\psi}(z,t_1), |0\rangle) - \xi(\hat{\psi}^{(0)}(z,t_0), |0\rangle) \tag{17}$$

To evaluate this we must determine the change in the field operator over time. The time evolution of the field operator in the presence of an electric field is given by (6). The solution, for time $t \geq t_0$ is,

$$\hat{\psi}(z,t) = \sum_r \left( \hat{b}_r \varphi_{1,r}(z,t) + \hat{d}_r^\dagger \varphi_{-1,r}(z,t) \right) \tag{18}$$

where the $\varphi_{\lambda,r}(z,t)$ evolve according to,

$$i \frac{\partial \varphi_{\lambda,r}(z,t)}{\partial t} = H \varphi_{\lambda,r}(z,t) \tag{19}$$

and where the $\varphi_{\lambda,r}(z,t)$ satisfy the initial condition $\varphi_{\lambda,r}(z,t_0) = \varphi_{\lambda,r}^{(0)}(z,t_0)$ so that the field operator $\hat{\psi}(z,t)$ will satisfy the initial condition $\hat{\psi}(z,t_0) = \hat{\psi}^{(0)}(z,t_0)$. Use (18) along with (9), (10), and (15) to obtain,

$$\xi\left(\hat{\psi}(z,t),|0\rangle\right) = \sum_r \int \varphi_{-1,r}^\dagger(z,t) H_0 \varphi_{-1,r}(z,t) dz \tag{20}$$

Therefore the change in the energy from $t_0$ to $t_1$ is given by,

$$\Delta \xi_{vac} = \sum_r \left( \begin{array}{c} \int \varphi_{-1,r}^\dagger(z,t_1) H_0 \varphi_{-1,r}(z,t_1) dz \\ -\int \varphi_{-1,r}^\dagger(z,t_0) H_0 \varphi_{-1,r}(z,t_0) dz \end{array} \right) \tag{21}$$

Use the initial condition $\varphi_{\lambda,r}(z,t_0) = \varphi_{\lambda,r}^{(0)}(z,t_0)$ along with (13) and (14) to obtain,

$$\Delta \xi_{vac} = \sum_r \delta \varepsilon_{-1,r} \tag{22}$$

where,

$$\delta \varepsilon_{\lambda,r} = \left( \int \varphi_{\lambda,r}^\dagger(z,t_1) H_0 \varphi_{\lambda,r}(z,t_1) dz \right) - \varepsilon_{\lambda,r}^{(0)} \tag{23}$$

### **4. Calculating the change in energy.**

We shall now apply the results of the last section to a specific perturbation. Let the electric potential $V(z,t)$ be given by,

$$V(z,t) = 4 \cos(k_w z) \left( \frac{\sin(mt)}{t} \right) \tag{24}$$

where m is the mass of the electron and $k_w = 2\pi w/L < m$ where w is a positive integer. It is obvious from the above expression that $V(z,t) \to 0$ at $t \to \pm\infty$. Under the action of this electric potential each initial wave function $\varphi_{\lambda,r}^{(0)}(z,t_0)$, where $t_0 \to -\infty$, evolves into the final wave function $\varphi_{\lambda,r}(z,t_1)$ where $t_1 \to +\infty$. We need to solve for $\delta\varepsilon_{\lambda,r}$.

This problem has already been addressed in [1] and a solution was obtained using standard perturbation theory. In [1] it was shown that $\delta\varepsilon_{\lambda,r}$ can be expressed as the following expansion in the parameter $q$,

$$\delta\varepsilon_{\lambda,r} = q\delta\varepsilon_{\lambda,r}^{(1)} + q^2\delta\varepsilon_{\lambda,r}^{(2)} + O(q^3) \tag{25}$$

where $O(q^3)$ means terms to the third order of $q$ and higher. For the problem at hand it was shown in [1] that,

$$\delta\varepsilon_{\lambda,r}^{(1)} = 0 \tag{26}$$

and,

$$\delta\varepsilon_{\lambda,r}^{(2)} = 2\pi^2 \lambda k_w \left( \frac{(p_r + k_w)}{E_{r+w}} - \frac{(p_r - k_w)}{E_{r-w}} \right) \tag{27}$$

It was also shown that $\delta\varepsilon_{-1,r}^{(2)} < 0$ for all $r$. Therefore in the limit of small $q$ we obtain,

$$\delta\varepsilon_{-1,r} = q^2\delta\varepsilon_{-1,r}^{(2)} + O(q^3) \underset{q \to 0}{=} q^2\delta\varepsilon_{-1,r}^{(2)} < 0 \tag{28}$$

Use this in (22) to obtain,

$$\Delta\xi_{vac} = \sum_r \delta\varepsilon_{-1,r} = \sum_r \left( q^2\delta\varepsilon_{-1,r}^{(2)} + O(q^3) \right) \underset{q \to 0}{=} q^2 \sum_r \delta\varepsilon_{-1,r}^{(2)} < 0 \tag{29}$$

It is shown in [1] that this equals,

$$\Delta\xi_{vac} \underset{q \to 0}{=} -4\pi q^2 k^2 L \tag{30}$$

As can be seen from this result the change in the energy is negative so that energy of the final state is less than the energy of the initial unperturbed vacuum state. Therefore energy has been extracted from the vacuum state due to the application of the electric field.

## 5. Conclusion.

We start out with a system in the initial unperturbed vacuum state $\left(\hat{\psi}^{(0)}(z,t_0), |0\rangle\right)$ at the initial time $t_0 \to -\infty$ where the electric potential is zero. Under the action of the electric potential of Eq. (24) the field operator evolves in time according to (6). The system is in the state $\left(\hat{\psi}(z,t_1),|0\rangle\right)$ at the final time $t_1 \to \infty$ where the electric potential is, once again, zero. The change in energy from the initial to the final state is then calculated. It is shown that, for sufficiently small $q$, this change is negative. Therefore the final state $\left(\hat{\psi}(z,t_1),|0\rangle\right)$ has less energy than that of the vacuum state. The conclusion is that in the Heisenberg picture of QFT states must exist with less energy than that of the vacuum state. The is the same result that was derived in Ref. [1] for Dirac's hole theory.

**References**

1. D. Solomon. Physc. Scr. **74** (2006) 117-122. (see also quant-ph/0607037)
2. W. Greiner, B. Muller, and J. Rafelski. Quantum electrodynamics of strong fields. Springer-Velag, Berlin, 1985.
3. W. Pauli. Pauli Lectures on Physics Vol 6 Selected Topics in Field Quantization, MIT Press, Cambridge, Massachusetts, 1973.